\documentclass{eptcs}
\providecommand{\event}{ICE 2010}
\usepackage[pdftex]{graphicx}
\usepackage{subfigure}
\usepackage{multicol}
\usepackage{amsmath}
\usepackage{amssymb}

\newcommand{\bb}{\approx_b}
\newcommand{\colonequals}{\mathrel{\mathrel{\mathop:}=}}

\newcommand{\Sys}{Sys}
\newcommand{\Com}{Comp}
\newcommand{\Act}{Act}
\newcommand{\Int}{Int}
\newcommand{\Beh}{Beh}
\newcommand{\port}[2]{\mathord{#1\mathord{:}#2}}
\newcommand{\ports}[1]{\operatorname{ports}(#1)}
\newcommand{\lts}[1]{\mathopen{[\mkern-2.67mu[}#1\mathclose{]\mkern-2.67mu]}}
\usepackage{ifthen}
\newcommand{\firstargument}{}
\newcommand{\tran}[1][\empty]{%
	\begingroup%
	\renewcommand{\firstargument}{#1}%
	\tranTwo}
\newcommand{\tranTwo}[2][\empty]{%
	\mathrel{%
		\stackrel{%
			\ifthenelse{\equal{#2}{\empty}}{\phantom{a}}{\makebox(0,0){$\scriptstyle#2$}}
			}{%
			\longrightarrow}%
			\ifthenelse{\equal{\firstargument}{\empty}\AND\equal{#1}{\empty}}{}{%
			{\!\!}_{%
					{\firstargument}
				}^{#1}%
			}%
		}%
	\endgroup}

\title{Port Protocols for Deadlock-Freedom of Component Systems}
\author{Christian Lambertz \qquad\qquad Mila Majster-Cederbaum
\institute{University of Mannheim, Germany}
\email{lambertz@informatik.uni-mannheim.de}
}
\def\titlerunning{Port Protocols for Deadlock-Freedom of Component-Based Systems}
\def\authorrunning{C. Lambertz \& M. Majster-Cederbaum}
\begin{document}
\maketitle

\begin{abstract}
In component-based development, approaches for property verification exist that avoid building the global system behavior of the component model.
Typically, these approaches rely on the analysis of the local behavior of fixed sized subsystems of components.
In our approach, we want to avoid not only the analysis of the global behavior but also of the local behaviors of the components. 
Instead, we consider very small parts of the local behaviors called port protocols that suffice to verify properties.
\end{abstract}

\section{Introduction}

Component-based development (CBD) helps to master the design complexity of software systems and enhances reusability. 
In formal CBD models, each component typically offers a set of ports for cooperation with other components.
Thereby, restrictions on the architecture of the system and the behavior of components allow to verify properties such as deadlock-freedom without exhaustively searching the global state space.
Here, we consider the CBD model of \emph{interaction systems} by Sifakis et.\ al.\ \cite{Sifakis03} in which data and I/O operations are completely abstracted away and every single operation is called an action.
Each component's behavior is modeled as a labeled transition system (LTS), where the set of labels equals the set of actions and each action is understood as a port of the associated component.
The actions are then grouped into sets called \emph{interactions} to model cooperation. 
Thereby, any action can only be executed if all other actions contained in an appropriate interaction are also executable. 
The global behavior is then derived by executing the interactions nondeterministically according to their executability.

A drawback of the original model is that a port is considered as a single action and thus no additional behavior can be specified for it.
Here, we extend the model of interaction systems to also capture port behavior.
We group several actions of one component and call this group a \emph{port alphabet}.
Additionally, every port alphabet is equipped with a LTS which we call \emph{port protocol}.
The idea behind this approach is that in verification steps we use the port protocols of involved components instead of their LTSs. 
This is more efficient since the behavior of the component is typically much larger (if we compare the number of states and transitions) than its port protocols. 
The verification of properties for the whole component then follows from the verification step that used only the port protocols.
Furthermore, this supports a gray box view of the components that is desired in CBD similar to the principle of information hiding~\cite{Bruin00}.

In Hennicker et.\ al.\ \cite{Hennicker08} and Mota et.\ al.\ \cite{Sampaio09}, we find similar ideas. 
In \cite{Hennicker08}, each port provides a protocol which is correct w.r.t.\ its component, i.e., the behavior of the component restricted to the actions in the port alphabet (i.e., any other action becomes unobservable) is weak bisimilar to the port protocol. 
Then, a notion called ``neutrality'' allows to apply a reduction strategy such that properties need only to be verified on the reduced part of the system.
Thereby, neutrality of a port $q$ for a port $p$ means that the composition of the protocols of $p$ and $q$ restricted to the alphabet of $p$ is weak bisimilar to the protocol of $p$, i.e., it is sufficient to consider only $p$.
In \cite{Sampaio09}, a similar idea is called ``compatibility'' of two ports and requires that all sequences of actions of one port are also possible in the other one.

After introducing our definitions in Sec.~2, we consider in Sec.~3 an example where two ports are neither neutral nor compatible, but our approach presented in Sec.~4 allows to verify its deadlock-freedom.

Note that several approaches for proving deadlock-freedom in interaction systems exist, e.g., Majster-Cederbaum and Martens \cite{Martens08} or Bensalem et.\ al.\ \cite{Bensalem09} in the context of BIP \cite{Sifakis06} (for which interaction systems are a theoretical model).
Apart from the lack of a gray box view of the components, which is desired in CBD~\cite{Bruin00}, these approaches also exploit the compositional structure of the system.
In Sec.~3, we demonstrate how the approach of \cite{Martens08} can further benefit from the introduction of port protocols by means of an example system.

The approach of \cite{Bensalem09} is based on finding invariants for the components, which must be provided for each property, and for the interactions, which are computed automatically.
Unfortunately, according to Bensalem et.\ al.\ \cite{Bensalem08}, for this computation ``there is a risk of explosion, if exhaustiveness of solutions is necessary in the analysis process.''
Thus, this approach is not guaranteed to be polynomial in the number and size of the components respectively the port protocols which is an important property of our approach.
However, the introduction of port protocols in BIP could be a promising extension w.r.t.\ a gray box view, if component invariants can be established from the information only available from the port protocols. 

\section{Formalization of Protocol Interaction Systems}

A \emph{protocol interaction system} is defined by a tuple $\Sys \colonequals (\Com, \{P_i\}_{i \in \Com}, \{A_i^p\}_{i \in \Com \wedge p \in P_i}, \Int, \Beh)$.
Here, $\Com$ is a finite set of \emph{components}, which are referred to as $i \in \Com$. 
The available \emph{ports} of a component $i$ are given by the finite set $P_i$, and the mapping $\ports{i} \colonequals \{\port{i}{p} \mid p \in P_i \}$ allows to refer to a port $p$ of $i$ as $\port{i}{p} \in \ports{i}$.
The \emph{actions} of each port $\port{i}{p}$ are given by the set $A_i^p$, also denoted by \emph{port alphabet} $A_{\port{i}{p}}$, and are assumed to be disjoint, i.e., $\forall \, i,j \in \Com, p \in P_i, q \in P_j\colon i \not=j \vee p \not= q \implies A_i^p \cap A_j^q = \emptyset$.
All available actions of a component $i$ are contained in the \emph{action set} $A_i \colonequals \bigcup_{p \in P_i} A_i^p$, and the union of all action sets is called the \emph{global action set} $\Act \colonequals \bigcup_{i \in \Com} A_i$.

A nonempty finite set $\alpha \subseteq \Act$ of actions is called an \emph{interaction}, if it contains at most one action of every component, i.e., $\vert \alpha \cap A_i \vert \le 1$ for all $i \in \Com$. 
For any interaction $\alpha$ and component $i$ we put $i(\alpha) \colonequals A_i \cap \alpha$. 
Similarly, for $\alpha$ and a port $\port{i}{p}$ of $i$ we put $\port{i}{p}(\alpha) \colonequals A_{\port{i}{p}} \cap \alpha$.
The \emph{interaction set} $\Int$ contains all available interactions and covers all actions, i.e., we require that $\bigcup_{\alpha \in \Int} \alpha = \Act$ holds.

The \emph{behavior model} $\Beh$ of $\Sys$ contains for every component $i$ a LTS $\lts{i} \colonequals (S_i, A_i, \{\mathord{\tran[i]{a}}\}_{a \in A_i}, I_i )$ describing the \emph{local behavior} of $i$ where $S_i$ is the \emph{local state space}, action set $A_i$ contains the labels, $\{\mathord{\tran[i]{a}}\}_{a \in A_i}$ is a family of \emph{transition relations} with $\mathord{\tran[i]{a}} \subseteq S_i \times S_i$, and $I_i \subseteq S_i$ is the set of \emph{local initial states}.
Whenever $(s,s^{\prime}) \in \mathord{\tran{a}}$ we write $s \tran{a} s^{\prime}$ instead.
For every port $\port{i}{p}$ of $i$, $\Beh$ contains a LTS $\lts{\port{i}{p}} \colonequals (S_{\port{i}{p}}, A_{\port{i}{p}} \cup \{\tau\}, \{\mathord{\tran[\port{i}{p}]{a}}\}_{a \in A_{\port{i}{p}} \cup \{\tau\}}, I_{\port{i}{p}})$ describing the \emph{port protocol} of $\port{i}{p}$.
The special symbol $\tau$ is used to model unobservable behavior, and we require $\tau \notin \Act$, i.e., no component uses $\tau$ as an action.
However, the port protocols are allowed to contain $\tau$-transitions.

A port $\port{i}{p}$ of a component $i$ is said to be \emph{conform} to the component if $\lts{\port{i}{p}} \bb \operatorname{f}_{\port{i}{p}}(\lts{i})$ where $\mathord{\bb}$ denotes branching bisimilarity~\cite{Glabbeek96} and $\operatorname{f}_{\port{i}{p}}(\cdot)$ is a relabeling function that replaces all labels respectively transitions not contained in the port alphabet $A_{\port{i}{p}}$ with the label $\tau$ respectively with a $\tau$-transition.
Thereby, we assume that the port protocols are minimized w.r.t.\ branching bisimilarity.

Note that we use branching bisimilarity instead of weak bisimilarity, which is used in the approach of  Hennicker~et~al.~\cite{Hennicker08}, because branching bisimilarity preserves more properties of systems (a logical characterization of $\mathord{\bb}$ in \mbox{CTL*--X} exists~\cite{Nicola95}), it is more efficient to calculate~\cite{Groote90}, and, as remarked by van~Glabbeek and Weijland~\cite{Glabbeek96}, many systems that are weak bisimilar are also branching bisimilar. 

In the following, we fix a protocol interaction system $\Sys$.
The \emph{global behavior} of $\Sys$ is a LTS $\lts{\Sys} \colonequals (S, \Int, \{\mathord{\tran{\alpha}}\}_{\alpha \in \Int}, I)$ where the set of \emph{global states} $S \colonequals \prod_{i \in \Com} S_i$ is given by the Cartesian product of the local state spaces, which we consider to be order independent.
Global states are denoted by tuples $s \colonequals (s_{1}^{},\dots,s_{n}^{})$ with $n = \vert \Com \vert$, and the set of \emph{global initial states} is $I \colonequals \prod_{i \in \Com} I_i$. 
The family of \emph{global transition relations} $\{\mathord{\tran{\alpha}}\}_{\alpha \in \Int}$ is defined canonically where for any $\alpha \in \Int$ and any $s,s^{\prime} \in S$ we have $s \tran{\alpha} s^{\prime}$ if $\forall \, i \in \Com \colon \text{if } i(\alpha) = \{a_i\}$ then $s_{i}^{} \tran[i]{a_i} s_{i}^{\prime}$ and if $i(\alpha) = \emptyset$ then $s_{i}^{} = s_{i}^{\prime}$.

Let $C \subseteq \Com$ be a set of components.
The \emph{partial behavior} of $\Sys$ with respect to $C$ is a LTS $\lts{C} \colonequals (S_C, \Int_C, \{\mathord{\tran[C]{\alpha}}\}_{\alpha \in \Int_C}, I_C)$ where $S_C \colonequals \prod_{i \in C} S_i$, $I_C \colonequals \prod_{i \in C} I_i$, $\Int_C \colonequals \{ \alpha \cap (\bigcup_{i \in C} A_i) \mid \alpha \in \Int\} \setminus \{\emptyset\}$, and $\{\mathord{\tran[C]{\alpha}}\}_{\alpha \in \Int_C}$ is defined analogously to the family of global transition relations.

Let $P \subseteq \bigcup_{i \in \Com} \ports{i}$ be a set of ports.
The \emph{port behavior} of $\Sys$ with respect to $P$ is a LTS $\lts{P} \colonequals (S_P, \Int_P \cup \{\tau\}, \{\mathord{\tran[P]{\alpha}}\}_{\alpha \in \Int_P \cup \{\tau\}}, I_P)$ where $S_P \colonequals \prod_{\port{i}{p} \in P} S_{\port{i}{p}}$, $I_P \colonequals \prod_{\port{i}{p} \in P} I_{\port{i}{p}}$, and $\Int_P \colonequals \{ \alpha \cap (\bigcup_{\port{i}{p} \in P} A_{\port{i}{p}}) \mid \alpha \in \Int\} \setminus \{\emptyset\}$. 
For any $\alpha \in \Int_P$ and any $s,s^{\prime} \in S_P$ we have $s \tran[P]{\alpha} s^{\prime}$ if $\forall \, \port{i}{p} \in P \colon \text{if } \port{i}{p}(\alpha) = \{a_{\port{i}{p}}\}$ then $s_{\port{i}{p}}^{} \tran[\port{i}{p}]{a_{\port{i}{p}}} s_{\port{i}{p}}^{\prime}$ and if $\port{i}{p}(\alpha) = \emptyset$ then $s_{\port{i}{p}}^{} = s_{\port{i}{p}}^{\prime}$.
Additionally, for $s, s^{\prime} \in S_P$ we have $s \tran[P]{\tau} s^{\prime}$ if $\exists \, \port{i}{p} \in P \colon s_{\port{i}{p}}^{} \tran[\port{i}{p}]{\tau} s_{\port{i}{p}}^{\prime}$ and $\forall \, \port{j}{q} \in P \setminus \{\port{i}{p}\} \colon s_{\port{j}{q}}^{} = s_{\port{j}{q}}^{\prime}$. 

Finally, define the \emph{protocol communication graph} $G \colonequals (V,E)$ of $\Sys$ where the vertices are given by $V \colonequals \Com \cup (\bigcup_{i \in \Com} \ports{i})$ and 
the edges by $E \colonequals \{ \{i, \port{i}{p}\} \mid i \in \Com \wedge \port{i}{p} \in \ports{i}\} \cup \{ \{\port{i}{p}, \port{j}{q}\} \mid i,j \in \Com \wedge \port{i}{p} \in \ports{i} \wedge \port{j}{q} \in \ports{j} \wedge \exists \, \alpha \in \Int\colon \port{i}{p}(\alpha) \not= \emptyset \wedge \port{j}{q}(\alpha) \not= \emptyset\}$.
Two ports are \emph{connected} if they are related by an edge in $G$.
The \emph{port connectivity} of a port $\port{i}{p}$ is defined as the number of ports to which $\port{i}{p}$ is connected.
If the port connectivity of a port is less than two, we say that the port is \emph{uniquely connected}.
If $G$ forms a tree in the graph-theoretical sense, we say that $G$ is \emph{tree-like}.

We call a LTS \emph{deadlock-free} if all its states, which are reachable from an initial state, have at least one outgoing transition.

\section{Two Example Systems}

We present two examples: The first one shows that the approaches of Hennicker et.\ al.\ \cite{Hennicker08} and Mota et.\ al.\ \cite{Sampaio09} are not always applicable and the second how deadlock analysis in interaction systems can benefit from port protocols.

\begin{figure}[h]
\centering
\subfigure[Prot. comm. graph]{%
\includegraphics{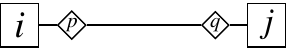}%
}
\hfill
%
\subfigure[Behavior and port protocol of $i$ resp.\ $\port{i}{p}$]{%
\includegraphics{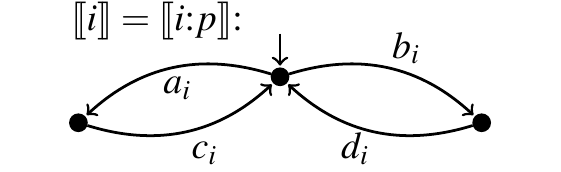}%
}
\hfill
\subfigure[Behavior and port protocol of $j$ resp.\ $\port{j}{q}$]{%
\includegraphics{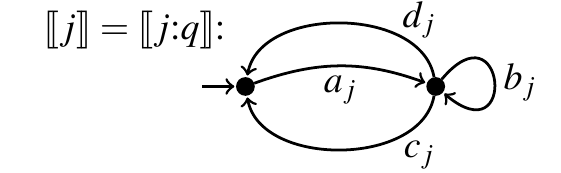}%
}
\caption{Protocol interaction system $Sys_{\text{ex1}}$ with $Int = \big\{ \{a_i, a_j\}, \{b_i, b_j\},\{c_i, c_j\},\{d_i, d_j\} \big\}$}
\label{fig:PIS_other}
\end{figure}

The first example is the protocol interaction system $Sys_{\text{ex1}}$ shown in Fig.\ \ref{fig:PIS_other} with $\Com = \{i,j\}$, $\ports{i} = \{\port{i}{p}\}$, and $\ports{j} = \{\port{j}{q}\}$.
The interaction set is given by $\Int = \big\{ \{a_i, a_j\}, \{b_i, b_j\},\{c_i, c_j\},$ $\{d_i, d_j\} \big\}$.
Obviously, all ports are conform to their corresponding component.
The example shows that the two connected ports are neither neutral nor compatible, since they restrict each other, i.e., in $\lts{\{\port{i}{p},\port{j}{q}\}}$ only the execution path ``$\{a_i,a_j\} \; \{c_i,c_j\}$'' is possible which restricts either port.

\begin{figure}[h]
\centering
\setlength{\columnsep}{-30pt}
\begin{multicols}{3}
\subfigure[Protocol communication graph]{%
\includegraphics{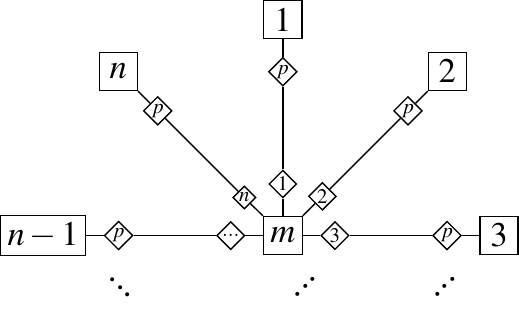}%
}

\columnbreak

\subfigure[Behavior of comp. $m$]{%
\includegraphics{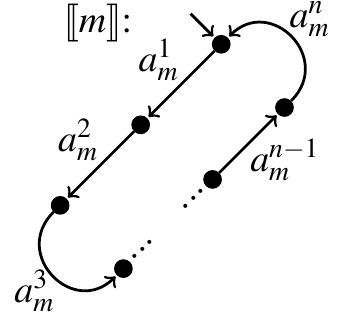}%
}

\columnbreak

\subfigure[Port prot.\ of port $\port{m}{i}$ with $1 \le i \le n$]{%
\includegraphics{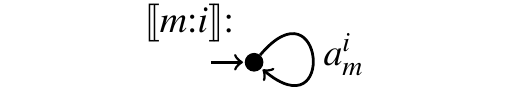}%
}
\subfigure[Behavior and port prot.\ of border comp.\ $i$ resp.\ port $\port{i}{p}$ with $1 \le i \le n$]{%
\includegraphics{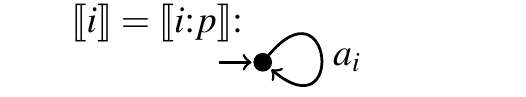}%
}
\end{multicols}
\caption{Protocol interaction system $Sys_{\text{ex2}}$ with $Int = \big\{ \{a_m^i, a_i\} \mid 1 \le i \le n \big\}$}
\label{fig:PIS_tree-like_motivator}
\end{figure}

The second example is the protocol interaction system $Sys_{\text{ex2}}$ shown in Fig.\ \ref{fig:PIS_tree-like_motivator} with $\Com = \{m,1,2,$ $\dots,n\}$, $\ports{m} = \{\port{m}{i} \mid i \in \Com \setminus \{m\} \}$, and $\ports{i} = \{\port{i}{p}\}$ for $i \in \Com \setminus \{m\}$.
The interaction set is given by $Int = \big\{ \{a_m^i, a_i\} \mid i \in \Com \setminus \{m\} \big\}$.
Obviously, all ports are conform to their corresponding component.

As an example for deadlock analysis, we consider the analysis of Majster-Cederbaum and Martens \cite{Martens08} for tree-like interaction systems. 
The check for deadlock-freedom of $Sys_{\text{ex2}}$ requires among other things that we analyze the partial behavior of all pairs of connected components, i.e., we have to carry out this analysis $n$ times.
Since the size of any such partial behavior is $\text{O}(n)$---because in each check the middle component is used---the whole deadlock analysis needs $\text{O}(n^2)$.
If we use the port protocols instead of the whole behavior in each step, each protocol behavior can be traversed in constant time. 
Thus, the total amount of work is $\text{O}(n)$.

Note that the example only motivates the use of port protocols in tree-like systems and that the global state space of the example can be traversed in $\text{O}(n)$. 
But, with more complex behavior of the non-middle components the cost of this traversal increases such that a global state space analysis becomes unfeasible.

\section{Proving Deadlock-Freedom}

In order to exploit the compositional information and the information obtained by combining the port protocols, we need to put restrictions on the architecture, e.g., on the form of the protocol communication graph, and on the local behaviors, e.g., on the existence of unobservable behavior in the port protocols.

The following theorem exploits such restrictions and allows for efficient verification of deadlock-freedom in interaction systems, which can be performed in time polynomial in the number and size of the port protocols.
Note that the examples of Sec.\ 3 can be verified in this way.

\vspace*{5pt}
\noindent\textbf{Theorem:} Let $\Sys$ be a protocol interaction system and $G$ its protocol communication graph.  
Assume that $G$ is tree-like and that every port is uniquely connected and conform to its corresponding component and its minimal port protocol w.r.t.\ branching bisimilarity is $\tau$-free.
If for all connected ports $\port{i}{p}$ and $\port{j}{q}$ of all components $i,j \in \Com$ holds that $\lts{\{ \port{i}{p}, \port{j}{q} \}}$ is deadlock-free then $\lts{\Sys}$ is deadlock-free.
\vspace*{5pt}

The theorem exploits the idea that an unobservable step in a port protocol is only present if the component's future behavior can be influenced by the cooperation with its environment.
If no $\tau$ is present, the component's behavior visible through the port protocol is inevitable.
Because of the structure of the protocol communication graph, it is then sufficient to check pairs of port protocols for deadlock-freedom, because due to their $\tau$- and deadlock-freedom, no cyclic waiting relation is possible.

\vspace*{5pt}
\noindent\textit{Proof (Sketch):} We successively consider partial behaviors of connected components of increasing size in an induction like manner.
Assume that there is a set $C \subseteq \Com$ of components such that $\lts{C}$ is deadlock-free. 
Now pick a component $j \notin C$ and consider $C' \colonequals C \cup \{j\}$.
Assume $\lts{C'}$ is not deadlock-free, although a component $i \in C$ and a port $\port{i}{p} \in \ports{i}$ exist such that $\port{i}{p}$ is connected to a port $\port{j}{q} \in \ports{j}$ and $\lts{\{\port{i}{p},\port{j}{q}\}}$ is $\tau$- and deadlock-free---which follows from the assumptions.
Due to the deadlock, there is a reachable state $s_{C'} \in S_{C'}$ that has no outgoing transition. 
Since the corresponding state $s_{C} \in S_{C}$ in the system without $j$ is deadlock-free, there must be an action $a_i \in A_i$ and states $s_i,s'_i \in S_i$ with $s_i$ being $i$'s local part in $s_{C}$ and $s_i \tran[i]{a_i} s'_i$.
But, the corresponding interaction $\alpha$ with $i(\alpha) = \port{i}{p}(\alpha) = \{a_i\}$ is not available in $s_{C'}$ anymore---due to the deadlock, i.e., there must be an $a_j \in A_j$ with $j(\alpha) = \port{j}{q}(\alpha) = \{a_j\}$ that is not enabled in the local part $s_j$ of $s_{C'}$.
 Consider the partial behavior $\lts{\{i,j\}}$ and the state $(s_i,s_j) \in S_{\{i,j\}}$, which is reachable from an initial state in $\lts{\{i,j\}}$ since the deadlocked state $s_{C'}$ is reachable in $\lts{C'}$.
Now, an equivalent state $(s_{\port{i}{p}},s_{\port{j}{q}}) \in S_{\{\port{i}{p},\port{j}{q}\}}$ with $s_i \bb s_{\port{i}{p}}$ and $s_j \bb s_{\port{j}{q}}$ is also reachable in the protocol behavior $\lts{\{\port{i}{p},\port{j}{q}\}}$ because of the protocol conformance.
But then, $\lts{C'}$ cannot be deadlocked since at least one $\alpha' \in \Int_{\{\port{i}{p},\port{j}{q}\}}$---and thus $\alpha' \in \Int$---is enabled in $(s_{\port{i}{p}},s_{\port{j}{q}})$ because of the protocol behavior's deadlock-freedom, and this $\alpha'$ can neither be blocked by $i$---because $i$'s cooperation with the other components in $C$ is deadlock-free---nor by $j$---because the only cooperation partner of $j$ is $i$---in $\lts{C'}$ because 
otherwise the protocol behavior $\lts{\{\port{i}{p},\port{j}{q}\}}$ would contain a $\tau$-transition. \hfill $\square$
\vspace*{5pt}

Currently, we are investigating weaker versions of the theorem, e.g., we conjecture that it is sufficient that the protocol behavior of combined port protocols is $\tau$-free instead of requiring the $\tau$-freedom of all port protocols, and we try to apply the port protocol approach to the verification of other generic properties such as progress and specific properties of a given system specified in \mbox{CTL*--X}.
Additionally, the proof of the theorem shows an application for a correctness-by-construction approach.


\end{document}